\documentclass[10pt, showpacs, prb, twocolumn, superscriptaddress]{revtex4}
\usepackage{graphicx}\usepackage{url}\usepackage{mathptmx}\usepackage{color}\usepackage{amssymb}
\begin{document}
\title{Aligned crystallite powder of NdFeAsO$_{0.86}$F$_{0.14}$: magnetic hysteresis and penetration depth}
\author{Yuri L. Zuev}
\affiliation{Materials Science and Technology Division, Oak Ridge National Laboratory, Oak Ridge, TN, 37831}
\author{Eliot D. Specht}
\affiliation{Materials Science and Technology Division, Oak Ridge National Laboratory, Oak Ridge, TN, 37831}
\author{Claudia Cantoni}
\affiliation{Materials Science and Technology Division, Oak Ridge National Laboratory, Oak Ridge, TN, 37831}
\author{David K. Christen}
\affiliation{Materials Science and Technology Division, Oak Ridge National Laboratory, Oak Ridge, TN, 37831}
\author{James R. Thompson}
\affiliation{Materials Science and Technology Division, Oak Ridge National Laboratory, Oak Ridge, TN, 37831}
\affiliation{Department of Physics, University of Tennessee, Knoxville, TN, 37996-1200}
\author{Rongying Jin}
\affiliation{Materials Science and Technology Division, Oak Ridge National Laboratory, Oak Ridge, TN, 37831}
\author{Athena S. Sefat}
\affiliation{Materials Science and Technology Division, Oak Ridge National Laboratory, Oak Ridge, TN, 37831}
\author{David G. Mandrus}
\affiliation{Materials Science and Technology Division, Oak Ridge National Laboratory, Oak Ridge, TN, 37831}
\author{Michael A. McGuire}
\affiliation{Materials Science and Technology Division, Oak Ridge National Laboratory, Oak Ridge, TN, 37831}
\author{Brian C. Sales}
\affiliation{Materials Science and Technology Division, Oak Ridge National Laboratory, Oak Ridge, TN, 37831}
\begin{abstract}
We report the basal-plane critical current and superfluid density of magnetically aligned NdFeAsO$_{0.86}$F$_{0.14}$ powder. This sample has individual crystallite grains permanently oriented with their $c$ axis along the external measuring field. Magnetic irreversibilities at high field suggest strong flux pinning of basal-plane critical currents, with monotonic field dependence and no evidence of the ``fishtail'' effect. The small particles provide a sensitive indicator of \textit{dc} flux penetration, and allow analysis of the temperature dependence of $ab-$plane London penetration depth $\lambda_{ab,\mathrm{L}}$, which is quadratic at low $T$. This feature may not necessarily be due to the nodes in the gap, but may be rather a sign of a strong pair-breaking.  A quantitative determination of the absolute magnitude of $\lambda_{ab,\mathrm{L}}$ is hindered by the need for accurate knowledge of the particle size distribution.  
\end{abstract}
\maketitle

\section{Introduction}
Discovery of a new superconducting family of FeAs-based superconductors has generated an explosion of research activity~\cite{Kamihara}. The superconductors in this family were produced by doping a parent compound, REFeAsO (``1111'' phase, RE-rare earth) with electrons by substitution of fluorine for oxygen, or by some oxygen deficiency. The highest $T_C$ obtained this way is 55 K in~\cite{Ren} SmFeAsO$_{0.9}$F$_{0.1}$. Hole doping by substituting strontium for rare earth also leads to superconductivity~\cite{Wen} with a $T_C=25$ K. Later, superconductivity in a related material BaFe$_2$As$_2$ (these are termed ``122'' materials) was discovered~\cite{Rotter}, where again, superconductivity occurs by doping either electrons (e.g. Co on Fe site, $T_C\approx 22$ K) or holes (e.g. K on Ba site, $T_C\approx 38$ K).  The parent (undoped) compounds of both 1111 and 122 materials are antiferromagnetic metals with a  spin-density wave setting in at around 130 K~\cite{Dong, Wang}. The Fermi surface consists of multiple sheets, electron- and hole-like in character~\cite{Singh}. Of the superconducting properties, most basic are the two Ginzburg-Landau length scales: the $ab$-plane values for the coherence length $\xi_{ab}$ is about 2-3 nm (e.g. Ref.~\cite{Jaroszynski}) and the magnetic penetration depth $\lambda_{ab}\approx 200$ nm (e.g. Ref.~\cite{Khasanov})

In this work we concentrate on NdFeAsO$_{0.86}$F$_{0.14}$ superconductor with a $T_C=48$ K, as a member of the iron pnictides 1111-family. Like other pnictide superconducting  compounds, it too has a layered, anisotropic crystal structure. Recently, the synthesis of small single crystals of such high $T_C$, 1111 iron pnictide superconductors has enabled characterization of some anisotropic superconducting properties~\cite{Jia, Jo}. It is important to further elucidate properties determined by strong basal-plane supercurrents. For such purposes, an alternative to small single crystals (typical mass $<1\;\mu$g~\cite{Jia, Jo, Zhigadlo}) is a large collection of small crystallites (here, total mass $\sim 10$ mg), whose orientation has been made to have a nearly common $c-$axis by magnetic alignment. Such samples actually have some advantages over small single crystals, since a sub-micron grain size allows higher sensitivity to small changes in magnetic penetration depth, while the large number of particles provides a strong signal. Therefore, it is possible to conduct meaningful measurements using low-frequency and $dc$ techniques.  In this article we report on making such a sample and  our findings from both the high field persistent (critical state) currents and the low field screening.

\section{Sample preparation and characterization}
Stoichiometric amounts of FeAs, NdF$_3$, Nd$_2$O$_3$, and Nd metal were ground together in a helium glove box, pressed into a pellet and sealed in a silica ampoule with 0.25 atmosphere of Ar gas. The pellet with nominal composition of NdFeAsO$_{0.86}$F$_{0.14}$ was heated to 1200$^{\circ}$ C for 15 h. Powder x-ray diffraction gave the correct phase with about 2-5\% of NdAs, FeAs, and NdOF impurities. The $T_C$ was 48 K, as determined by the low-field $ac$ and $dc$ susceptibility.

\begin{figure*}[t]
\centering
\includegraphics[width=\textwidth]{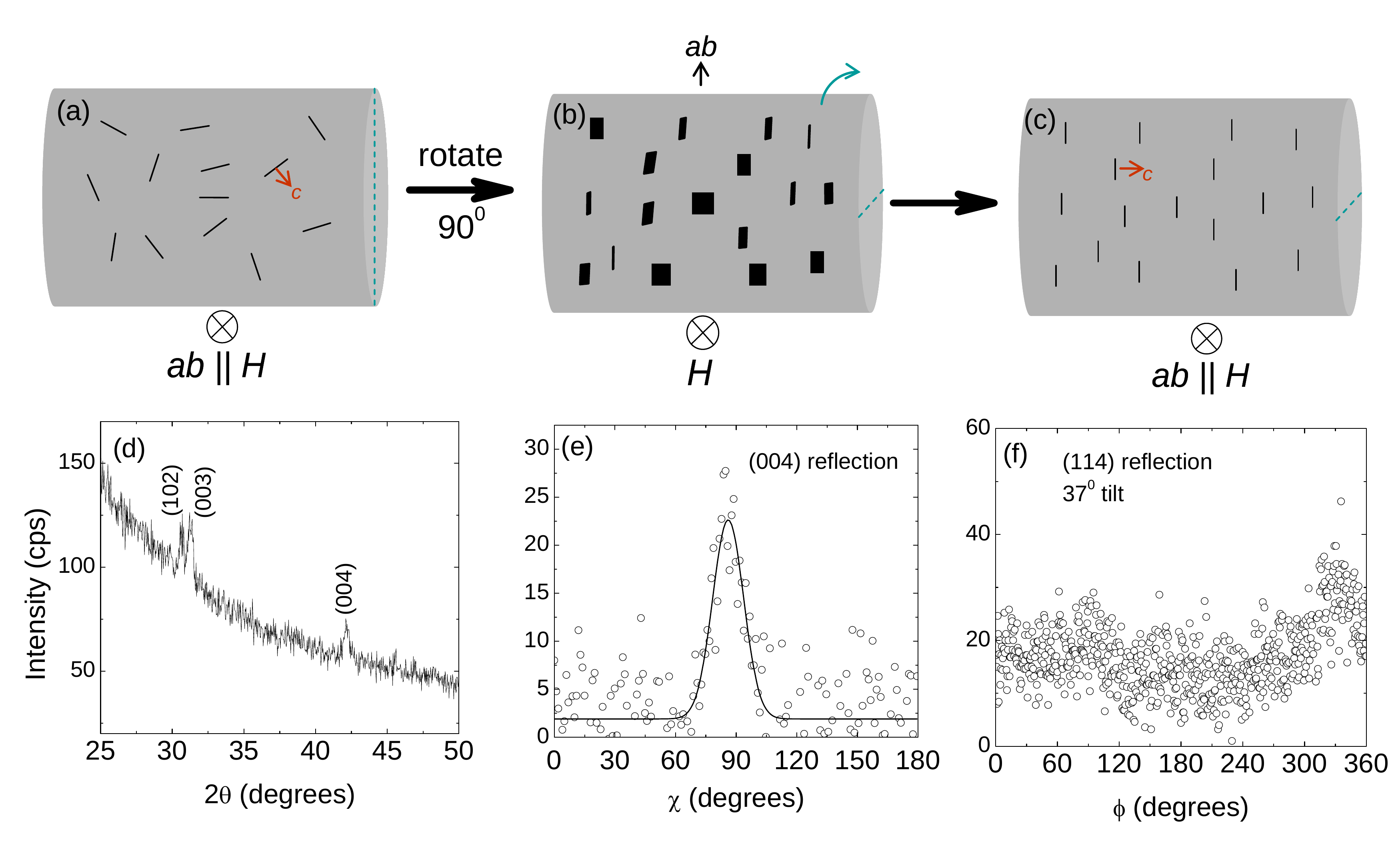}
\caption{(color online) Two-step alignment of the powder particles, shown as $a-b$ platelets for illustration only. Magnetic field of 6 T points into the page. The first step produces particles with $ab$ plane parallel to external field and their $c$-axis randomly oriented in a plane, perpendicular to the field (a). After 90$^{\circ}$ rotation (b), the second step aligns $c$  with the sample's long axis (c); (d) $\theta-2\theta$ scan shows mostly (00L) peaks and predominant $c$-axis orientation. Background is due to glass-like epoxy. (e) Rocking curve of (004) peak shows 18$^{\circ}$ FWHM spread in individual particles orientation. (f) Absence of 4-fold structure in $\phi$ scan of (114) reflection indicates no azimuthal ordering of powder particles}
\label{fig:XRay}
\end{figure*}
\begin{figure}[b]
\centering
\includegraphics[width=0.9\columnwidth]{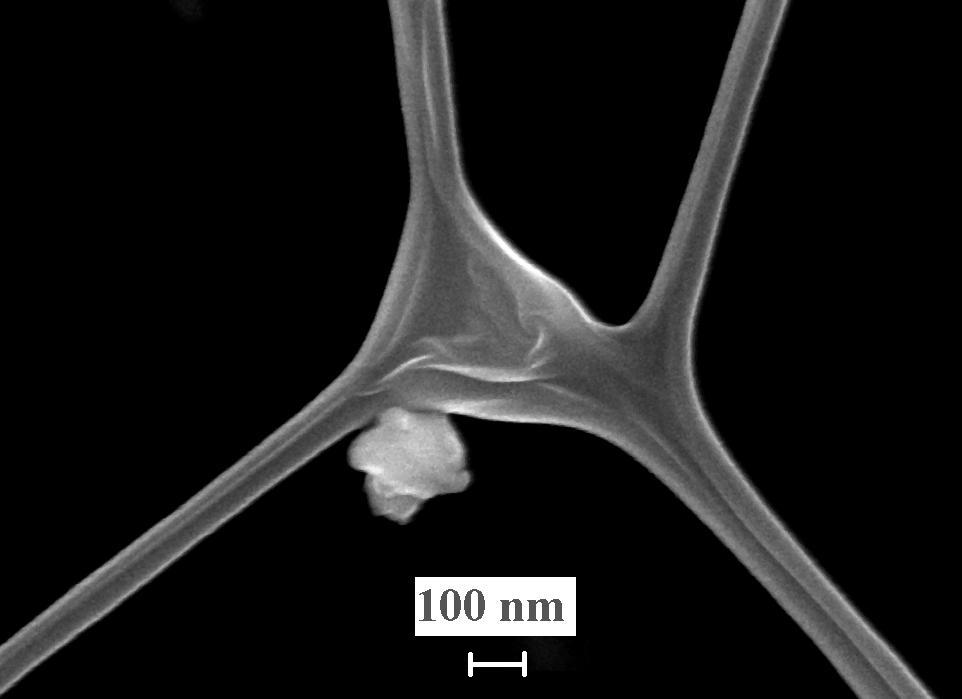}
\caption{SEM micrograph of a single particle on a holey carbon TEM grid, demonstrating diameter $\sim 200$nm and approximately equiaxial shape}
\label{fig:particle}
\end{figure}
A polycrystalline piece of NdFeAsO$_{0.86}$F$_{0.14}$ was mechanically ground to a fine powder, such that each particle constitutes a single grain. We did not use any sieve to remove large particles. A ten mg quantity of the as-ground powder was mixed with clear epoxy (Hardman 04004, setting time 1 hour) in a cylindrical gelatin capsule and placed in a magnetic field of 6 T, which  aligns the powder particles with $ab$ planes parallel to the field. A similar phenomenon was noted by H.-J. Grafe \textit{et al.}~\cite{Grafe} and B. C. Chang \textit{et al.}~\cite{Chang} although in those cases the rare earth ions were respectively spinless La$^{3+}$, and Sm$^{3+}$ with a magnetic moment of $\sim 0.7\mu_B$, while in our case it is Nd$^{3+}$ ($^4I_{9/2}$ ground state, which has a magnetic moment of $3.3\mu_B$ as a free ion). Therefore such an alignment cannot be ascribed to rare-earth magnetism, but rather results from layered, anisotropic crystal structure. We have performed the alignment procedure at room temperature in two steps to obtain $c$-axis orientation, as illustrated in Fig.~\ref{fig:XRay}. In the first step, field was applied perpendicular to the capsule axis, which produced a sample with $c$ axis randomly distributed  in the plane perpendicular to the applied field  (Fig.~\ref{fig:XRay}(a)). After about half hour, the capsule was rotated $90^{\circ}$ about its long axis (fig.~\ref{fig:XRay}(b)) and the epoxy was allowed to set. As a result, particles reoriented to align with a common $c$-axis (Fig.~\ref{fig:XRay}(c)), with a possible exception of a very small number of particles, whose $c$-axis was perpendicular to the capsule axis after the first alignment step. The total powder volume was $\sim 1.3$ mm$^3$ and the approximate capsule volume of 100 mm$^3$ resulted in a dilute $<1.5$ vol. \% particle composite dispersed  in the epoxy matrix.

Grain alignment was confirmed using a 4-circle x-ray diffractometer with a Cu K$\alpha$ source (50 kV, 100 mA), a sagitally focusing graphite monochromator, soller slits, and a NaI scintillation detector. A $\theta-2\theta$ scan (Fig.~\ref{fig:XRay}(d)) shows that the (003) and (004) reflections are enhanced relative to a random powder and the rocking curve for the (004) reflection is about 18$^{\circ}$ FWHM (Fig.~\ref{fig:XRay}(e)). This orientation spread is acceptable for studies of in-plane magnetic properties with field along the $c$ axis, but is too large if one wishes to study the behavior in parallel field. Only the $H\parallel c$ situation is examined in this paper. An azimuthal ($\phi$) scan at a 37$^{\circ}$ tilt (Fig.~\ref{fig:XRay}(f)) shows that the (114) intensity has a weak 2-fold variation which we attribute to the sample shape, i.e. there is no indication of an azimuthal grain alignment, which would produce at least a 4-fold symmetry.


\begin{figure}[t]
\centering
\includegraphics[width=\columnwidth]{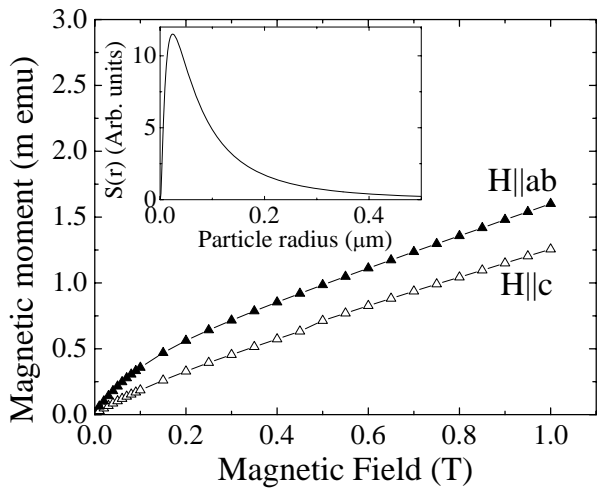}
\caption{Magnetic moment of aligned particles in epoxy for field applied $\parallel ab$ plane (filled symbols) and $\parallel c$ axis (open symbols) at room temperature. Particles align with a larger magnetic moment along the applied field. Inset: grain size distribution function, obtained by light scattering from a sample of the powder.}
\label{fig:anisotropy}
\end{figure}

For a quantitative analysis of magnetic data on such samples it is necessary to know the particle size distribution. An SEM image of a particle on a holey carbon TEM sample grid (Fig.~\ref{fig:particle}) shows a diameter of approximately 200 nm and close to an equiaxed particle shape. More detailed information about particle sizes was obtained using a NICOMP ZLS particle sizing system~\cite{nicomp}, which deduces particle size distribution from temporal correlations in intensity of light scattered off the particles diffusing in a liquid of a known viscosity and refraction index (we used diffusion pump fluid 704). The output, a smooth log-normal distribution $S(r)$ (Fig.~\ref{fig:anisotropy} inset),  shows that most of the particles are in sub-micron range. In a small applied field, superconducting particles smaller than a magnetic penetration depth produce little diamagnetism. Rather, the particles in the tail of the distribution, in the range of about a micron and above, most affect the measurement. Therefore the knowledge of the distribution tail is vital for quantitative analysis. 

Following the sample preparation we have confirmed that at room temperature the magnetization was higher for field along the $ab$ plane, than along the $c$ axis, as shown in fig.~\ref{fig:anisotropy}. The ratio of $M_{ab}/M_c=1.26$ at $H=1$ T is similar to that obtained by Chang \textit{et al.}~\cite{Chang}. Particles align with a larger moment along the field. 

\section{Measurements}
\subsection{Magnetic hysteresis}
\begin{figure*}[t]
\centering
\includegraphics[width=\textwidth]{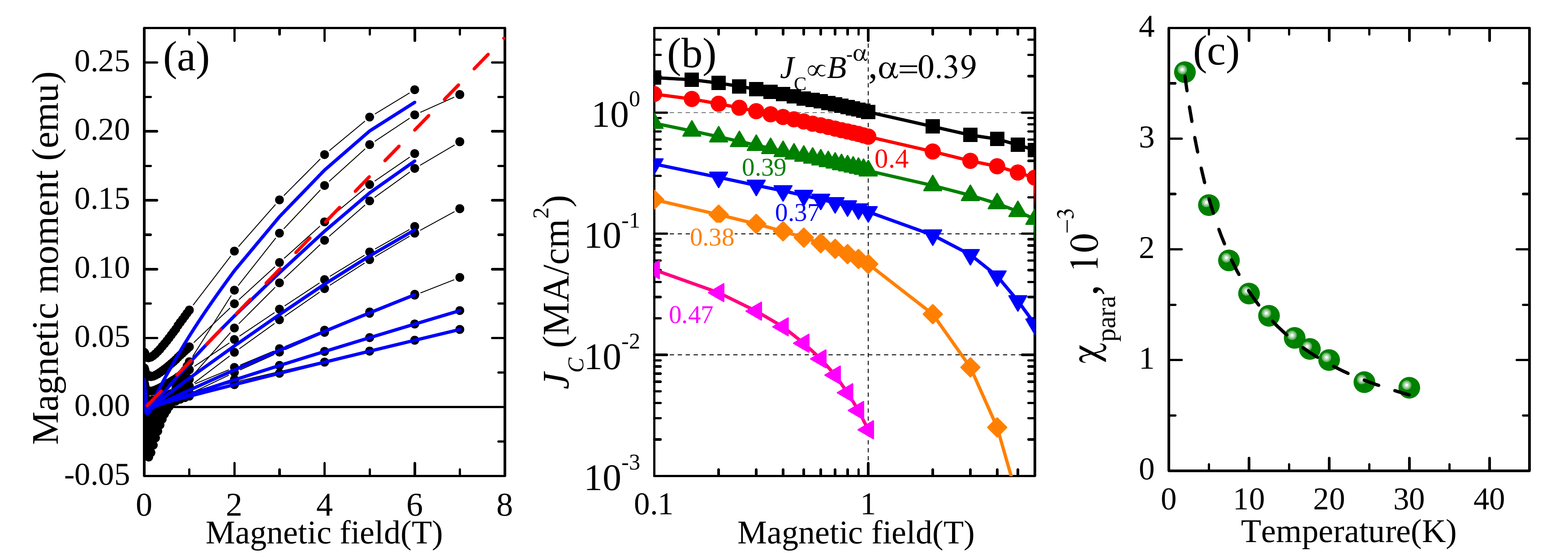}
\caption{(color online) (a) Magnetic hysteresis of the aligned powder samples at (top to bottom) $T=$ 2, 5, 10, 20, 30, and 40 K. Only RSO data are shown. Solid blue lines at the average of every hysteresis loop are an approximation for equilibrium magnetization $M_{eq}$. The slope of the dashed red line (shown only for $T=5$ K) serves as an estimate of paramagnetic susceptibility of Nd ions. (b) Critical current density, extracted from magnetic hysteresis data using Bean model for spherical particles (Eq.~\ref{eq:num}) for (top to bottom) $T=$ 2, 5, 10, 20, 30, and 40 K. Power law dependence of $J_C$ on field is indicated. (c) The temperature dependence of the background paramagnetism, given (for $T=5$ K) by the slope of the red dashed line in the panel (a), fits well with the Curie-Weiss law (black dashed line), producing $\theta=4.7K$ and an effective Nd moment of $2\mu_B$.}
\label{fig:hyst}
\end{figure*}
The magnetic hysteresis has been measured in \textit{dc} and oscillating sample (RSO) modes of the Quantum Design MPMS magnetometer. The agreement between the two methods is good. The  Fig.~\ref{fig:hyst}(a) shows the RSO data at several temperatures.

The $ab$-plane critical current density was extracted from the magnetic hysteresis measurements below 6 Tesla. Knowledge of particle sizes is crucial to quantitative analysis. Spherical particles were assumed with size distribution of the Fig.~\ref{fig:anisotropy} inset. Clem and Kogan~\cite{ClemKogan} found that the magnetic moment $m$ of a sphere of the radius $R$ and critical current density $J_C$ in a Bean critical state is given by
\begin{equation}
m=\frac{3\pi}{320}\frac{4}{3}\pi R^3 J_CR
\label{eq:msphere}
\end{equation}
Averaging with respect to particle size modifies the \emph{rhs} of the above equation:
\begin{equation}
M=\frac{m}{V}=\frac{3\pi}{320}J_C\frac{\int_0^\infty S(r)r^4\:dr}{\int_0^\infty S(r)r^3\:dr}
\label{eq:num}
\end{equation}
where $S(r)$ is the particle size distribution function of Fig.~\ref{fig:anisotropy} inset.

The extracted $J_C$ are shown in Fig.~\ref{fig:hyst}(b). Apparently, values for the in-plane critical current are relatively high. For example, at $T=5$ K and applied field $B_{\mathrm{app}}=6$ T the critical current density is about 0.3 MA/cm$^2$. It should be mentioned, that the magnitude of the $J_C$ is very sensitive to small uncertainties in the size distribution. The relative values are, of course, quite well defined.  

 The magnetic field dependence in Fig.~\ref{fig:hyst}(b) exhibits a distinct region of power-law behavior, $J_C\propto H^{-\alpha}$ with a nearly temperature-independent value for $\alpha$ of about 0.38  except near $T_C$, at 40 K. A weak dependence of $\alpha$ on $T$ has been found in high-$T_C$ cuprate coatings as well~\cite{ZuevAPL}. Unlike other reports in the literature on the irreversible properties of iron-arsenides~\cite{Zhigadlo, Yamamoto09, Yang}, we do not observe any ``peak'' or ``fishtail'' effects in $J_C(H)$.
 
 Figure~\ref{fig:hyst}(c) shows the temperature dependence of the background paramagnetism, which is predominantly due to Nd ions. This magnetic background has been estimated from the hysteresis loops of Fig.~\ref{fig:hyst}(a) as follows: first, a mean of the upper and lower hysteresis branch is calculated for each loop, as a proxy to the equilibrium magnetization $ M_{eq}\approx 1/2 (M_++M_-)$. These are shown in Fig.\ref{fig:hyst}(a) as solid blue lines. Then, a tangent to such center line, passing through origin, is constructed (for $T=5$ K this is shown in Fig.\ref{fig:hyst}(a) by the red dashed line). The slope of this tangent provides a good approximation for the background susceptibility $\chi_{\mathrm{para}}$. The Curie-Weiss fit to these values, $\chi_{\mathrm{para}}\propto (T+\theta)^{-1}$ (dashed line in Fig.~\ref{fig:hyst}(c)) produces a positive $\theta=4.7$ K, pointing to antiferromagnetic coupling between Nd ions and an effective Nd magnetic moment of about 2$\mu_B$, i.e. about 60\% of the free Nd$^{3+}$ ion value.  For $T>T_C$ the magnetic susceptibility decreased with increasing $T$, but could not be accurately described by a Curie-Weiss dependence. We attribute this, as well as the reduction in the inferred Nd$^{3+}$ magnetic moment value compared to that of a free ion to crystal-field effects in the aligned array of particles. Knowledge of the background paramagnetism will be required below to obtain a London penetration depth from the measured $ac$ magnetization. 

\subsection{Magnetic penetration depth}
\begin{figure}[b]
\centering
\includegraphics[width=\columnwidth]{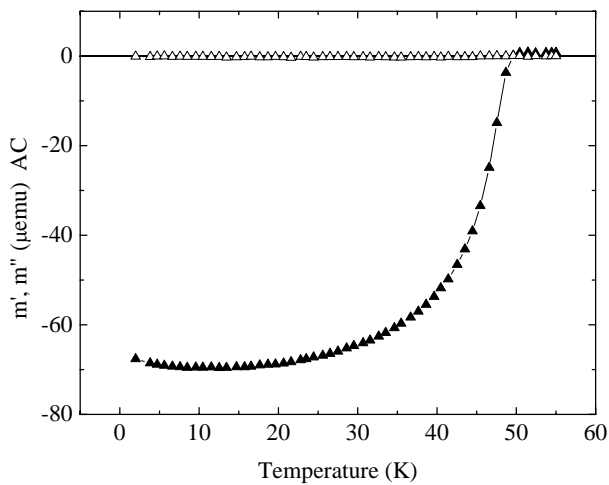}
\caption{ac magnetic moment of the aligned powder. Drive amplitude 1 Oe, frequency 1 kHz. $\blacktriangle$-in-phase component $m'$, $\triangle$-out-of-phase component $m''$. There is no peak in $m''$ because most particles are small compared to magnetic penetration depth.}
\label{fig:mT}
\end{figure}
We measured the low-field magnetization $M(T)$, using an \textit{ac} magnetometer (Quantum Design MPMS) between $T_C$ and 2 K in zero applied dc field (Fig.~\ref{fig:mT}). The drive amplitude and frequency were $H_{ac}=1$ Oe and $f=1$ kHz respectively. The epoxy had a weak, temperature-independent diamagnetism, about 0.2\% that of the sample's signal at low temperature. This background has been subtracted from the data. Knowing the particle size distribution $S(r)$, from the measured $M(T)$ we can extract the magnetic penetration depth $\lambda_{ab}$, using the equation 
\begin{equation}
4\pi M=-\frac{H_{ac}}{1-D}\frac{\int_0^\infty r^3 P(r/\lambda_{ab}) S(r)dr}{\int_0^\infty r^3 S(r) dr}
\label{eq:lambda}
\end{equation}
where, assuming spherical particles, we have a fraction of the grain's volume from which magnetic flux is excluded~\cite{Schoenberg} $P(r/\lambda_{ab})=[1-(3\lambda_{ab}/r)\coth(r/\lambda_{ab})+3\lambda_{ab}^2/r^2]$ and the demagnetizing ratio $D=1/3$. This is a straightforward technique, which generally agrees very well with other methods~\cite{Zuev}. However, as mentioned above, the relatively few large particles produce most of the diamagnetic signal. Unfortunately, the large $r$ tail of the size distribution is not known with enough accuracy to obtain reliable determination of the absolute value of the penetration depth. In contrast, the temperature dependence of $\lambda_{ab}$ is far less sensitive to uncertainties in size distribution, as shown below, so that we can obtain valuable results for the variation of $\lambda_{ab}$ with temperature.

\begin{figure}[t]
\centering
\includegraphics[width=\columnwidth]{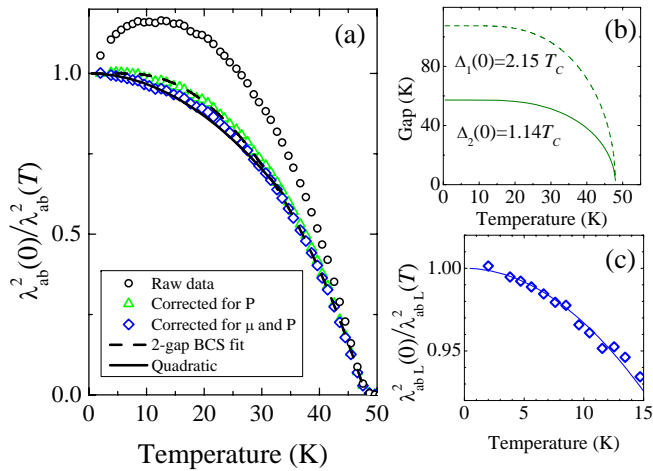}
\caption{(color online) (a) The $\lambda_{ab}^{-2}(T)$ extracted from the measured susceptibility $\chi(T)$. {\Large $\mathbf{\circ}$}- without corrections, \textcolor{green}{$\mathbf{\triangle}$}-corrected for extra signal from field-penetrated part, \textcolor{blue}{$\mathbf{\lozenge}$}-additionally corrected for magnetic permeability. Solid and dashed lines show quadratic and 2-gap BCS fits, respectively(b) The $T-$dependence of the superconducting gaps, used in two-gap BCS fit of \textcolor{green}{$\mathbf{\triangle}$}. (c) Enlarged low-temperature region, showing quadratic temperature dependence of the $\lambda_{ab\:\mathrm{L}}^{-2}$}
\label{fig:lambda}
\end{figure}

Circles in Fig.~\ref{fig:lambda}(a) show the $\lambda_{ab}^{-2}(T)$ extracted from the as-measured magnetization. There is a maximum near 10 K, below which the diamagnetism is reduced by the underlying Nd ion paramagnetism, as is evident also in Fig.~\ref{fig:mT}. Clearly, the Nd magnetic signal must be accounted for. Knowing the paramagnetic background from the previous section (Fig.~\ref{fig:hyst}(c)), we can do such a correction. We demonstrate here that, while it is possible to under-correct for the magnetic background and make the measurements appear to be consistent with a fully gapped state (with two different gaps on different Fermi sheets), a proper treatment of the background points toward an unconventional pairing or possibly a strong pair-breaking.

The Nd ion paramagnetism provides two contributions to the overall signal, as discussed by J. R. Cooper~\cite{Cooper}.  First, the paramagnetic signal comes from the grain's field-penetrated part, whose volume fraction is given by $1-P(r/\lambda)$. Following Ref.~\cite{Christen}, to subtract this paramagnetic signal we must replace $P(r/\lambda)$ in Eq.~(\ref{eq:lambda}) by  $P_{\mathrm{eff}}=P-4\pi\chi_{\mathrm{para}}(1-P)$. After such correction, the data of Fig.~\ref{fig:lambda}(a) show no maximum at any temperature (green triangles), but rather flat behavior at low temperatures, consistent with a fully-gapped state. On closer inspection, no simple $s$-wave BCS calculation fits the corrected data satisfactorily. Instead, as shown in Fig.~\ref{fig:lambda}(a) by the dashed black line, we obtain a good fit by assuming two gaps, with $T=0$ values $2.15\:k_BT_C$ and $1.14\:k_BT_C$, the smaller gap producing a quarter of the total superfluid fraction. To obtain the gap temperature dependence (Fig.~\ref{fig:lambda}(b)), the self-consistent BCS equation was solved numerically. As mentioned above, the shape of the $\lambda^{-2}(T)$ curve depends only weakly  on how well the size distribution is known. For reference, a change in particle size distribution that would decrease $\lambda^{-2}(0)$ by about a factor of 100, changes the large and small gap magnitudes to 2 and 1.23 of $k_BT_C$ respectively, and the relative weight of the smaller gap increases to about 35\%. Therefore, based on the correction $P\rightarrow P_{\mathrm{eff}}$ alone it could be concluded that this superconductor is fully gapped and is described well by the clean-limit BCS theory with two different gaps. Indeed, many authors have arrived at such a conclusion, as discussed below.

There is a second correction, however, that must be made in order to fully account for the magnetic background, and this one comes from the magnetic permeability in the London equation, $m\nabla\times\nabla\times\mathbf{B}=-\mu_0ne^2\mathbf{B}$, which describes the shielding of the magnetic field in the interior of a superconductor~\cite{Cooper}. For a magnetic material like NdFeAsOF, $\mu_0$ must be replaced with $\mu\mu_0=(1+4\pi\chi_{\mathrm{para}})\mu_0$ (factor of $4\pi$ is because the latter formula is in SI units, while values of fig.~\ref{fig:hyst}(c) are calculated in Gaussian units, and there is a difference of $4\pi$ between SI and Gaussian values of magnetic susceptibility). Then $\lambda^{-2}=\mu_0ne^2/m$ must be multiplied by $1+4\pi\chi_{\mathrm{para}}$ to obtain the true London penetration depth $\lambda_{ab,\mathrm{L}}$, which is the measure of the density of the superconducting condensate. This is shown by blue diamonds in the Fig.~\ref{fig:lambda}(a) and (c). At low temperatures these are best fit with a parabola $1-(T/55\mathrm{K})^2$ (solid blue line, Fig.~\ref{fig:lambda}(c)). Similar analysis and result can be found in Ref.~\cite{Martin}, from which we have gleaned the idea of this second correction. In that work, the role of $P_{\mathrm{eff}}$ is negligible due to large size of the sample, most of which is shielded from the magnetic field, while the correction for $\mu$ is large due to a much lower value of the Curie-Weiss $\theta=0.2$ K, treated as a fit parameter there, resulting in a much stronger $T$-dependence of $\chi_{\mathrm{para}}$ near $T=0$ compared with the present work where $\theta=4.7$ K has been obtained independently.

\section{Discussion}

There have been several reports on the temperature dependence of the superfluid density in various 1111 oxy-pnictides. From experimental work one can note measurements of H. Luetkens {\em et al.} (muons, LaFeAsO$_{1-x}$F$_x$)~\cite{Luetkens}, R. Khasanov {\em et al.} (muons, SmFeAsO$_{1-x}$F$_x$, NdFeAsO$_{1-x}$F$_x$)~\cite{Khasanov} on polycrystals, and L. Malone {\em et al.} (rf resonator, SmFeAsO$_{1-x}$F$_x$)~\cite{Malone}, K. Hashimoto {\em et al.} (microwave cavity, PrFeAsO$_{1-x}$F$_x$)~\cite{Hashimoto}, and C. Martin {\em et al.} (rf resonator, NdFeAsO$_{1-x}$F$_x$, LaFeAsO$_{1-x}$F$_x$)~\cite{Martin}, done on single crystalline samples. With the exception of Ref.~\cite{Martin} there is a common agreement that the superconducting state is fully gapped and can be accounted for by two gaps on different sheets of the Fermi surface. This manifests itself in temperature-independent $\lambda^{-2}$ at low temperatures, followed by convex temperature dependence. It is possible that this result would be different, had the proper corrections been applied. On the theoretical front, L. Benfatto {\em et al.}~\cite{Benfatto} have developed a model, where the superconducting properties are determined by an interband coupling between one electron- and one hole-like bands.  The model reproduces the main features of $\lambda^2(0)/\lambda^2(T)$, and agrees with the green dashed curve of Fig.~\ref{fig:lambda} quantitatively: $\Delta_1(0)=2T_C$ and $\Delta_2(0)=T_C$. The experiments of L. Malone {\em et al.}~\cite{Malone} provide similar estimates of two gap magnitudes. C. Martin \textit{et al.}~\cite{Martin}  also argue in favor of two gaps on different Fermi surface sheets, but find a $T^2$ behavior of $\lambda_{ab}(T)$ in single crystals of both NdFeAsO$_{1-x}$F$_x$ and LaFeAsO$_{1-x}$F$_x$ (where for the latter the paramagnetic background is absent and interpretation is more straightforward). Their findings are similar to the present result (blue diamonds, Fig.~\ref{fig:lambda}(a) and (c)) suggesting a possibility of unconventional superconductivity.  We would caution that in and of itself, the quadratic temperature dependence of the superfluid density  can not be taken as a sure sign of unconventional superconductivity. For example, in Nb/Ni bilayers the superfluid density displays a quadratic $T$-dependence due to pair-breaking proximity effect between superconducting Nb and ferromagnetic Ni (Ref.~\cite{TRL}). With the proposed $s^{\pm}$ pairing state of iron-pnictides~\cite{Mazin} the roles of magnetic and non-magnetic scatterers are expected to be interchanged: non-magnetic scattering will broaden the peak in the superconducting density of states until eventually one or both gaps close (perhaps not everywhere on the Fermi surface at once), while magnetic scattering will tend to make gaps equal in magnitude and reduce gap anisotropy by connecting states from different parts of the Fermi surface (Anderson theorem). Nagai et al.~\cite{Nagai} have analyzed a case of anisotropic gap on the electron-like Fermi surface (around the $M$ point of the Brillouin zone) and have shown, in particular, that the observed power-law temperature behavior of the spin-lattice relaxation rate $T_1^{-1}$(Ref.~\cite{Nakai}), considered by many to be at odds with a fully gapped state, finds a plausible explanation in that approach: while the density of states is fully gapped ($N=0$ for $|\epsilon|<\Delta_1$), it varies linearly with energy in the interval $\Delta_1\leq|\epsilon|\leq\Delta_2$, a feature reminiscent of the $d$-wave situation in cuprates. We note that a qualitatively similar behavior of $N(\epsilon)$, i.e. a finite sub-gap quasiparticle density of states could result from non-magnetic scattering. Overall, the question of the pairing state in iron oxypnictides is far from resolution at this stage. 

There have been several reports on the critical current densities of 1111 materials, all extracted from magnetic hysteresis of polycrystalline samples by means of the Bean model~\cite{Senatore,Chen}. The shortcoming of this approach is that the spatial scale of supercurrent circulation is often unknown: it can be given by an individual grain size (probably the case near $T_C$ and/or in high fields), or by the sample size, which is likely at low temperatures and fields in well connected materials~\cite{Yamamoto08}. The resulting inferred critical current density can differ by two orders of magnitude. In a single crystal of the 122 material Ba$_{1-x}$K$_x$Fe$_2$As$_2$, the in-plane critical current density reaches~\cite{Yang} close to $5\cdot 10^6$ A/cm$^2$ at 2 K, 0 T, a value comparable  to our result. 

\section{Conclusions}
We have obtained the temperature dependence of the $ab-$plane superfluid density $\lambda_{ab,\mathrm{L}}^{-2}(T)$ and critical current $J_C(T, H)$ as a function of temperature and magnetic field for an ensemble of aligned, sub-micron size grains of NdFeAsO$_{0.86}$F$_{0.14}$. The temperature dependence of the superfluid density, after proper corrections for background paramagnetism, is quadratic, which alone does not enable us to come to a conclusion regarding the nature of the pairing state.  The critical current density is high, which makes this superconductor potentially interesting for applications, such as superconducting wires. We have not observed any ``fishtail'' anomalies in $J_C(H)$.

We are indebted to Dr. Zheng Gai for use and help with the AC magnetometer, and Dr. Parans M. Paranthaman for providing access to the particle sizing instrument. Research sponsored by the Division of Materials Sciences and Engineering, Office of Basic Energy Sciences, U.S. Dept. of Energy. The research at Oak Ridge National Laboratory's Center for Nanophase Materials Sciences was sponsored by the Scientific User Facilities Division, Office of Basic Energy Sciences, U.S. Department of Energy. YLZ acknowledges support from the Oak Ridge Institute for Science and Education and the DOE Office of Electricity Delivery and Energy Reliability, Superconductivity Program for Electric Power Systems.

\end{document}